\documentclass[preprint]{revtex4}
\usepackage[dvips]{graphicx}
\usepackage{color}
\usepackage{amsmath}
\usepackage{amssymb}
\usepackage{bm}
\usepackage{comment}
\usepackage{textcomp}

\begin{document}

\title{Short time dynamics of tracer in ideal gas}

\author{Fumiaki Nakai\textsuperscript{1}}
\author{Yuichi Masubuchi\textsuperscript{2}}
\author{Takashi Uneyama\textsuperscript{2}}
\affiliation{\textsuperscript{1} Department of Materials Physics, Graduate School of Engineering, Nagoya University,
Furo-cho, Chikusa, Nagoya 464-8603, Japan}
\affiliation{\textsuperscript{2} Center for Computational Science, Graduate School of Engineering, Nagoya University,
Furo-cho, Chikusa, Nagoya 464-8603, Japan}

\begin{abstract}
A small tagged particle immersed in a fluid exhibits the Brownian motion
and diffuses at the long-time scale.
Meanwhile, at the short-time scale, the
dynamics of the tagged particle cannot be simply described by
the usual generalized Langevin equation with the Gaussian noise, since the number
of collisions between the tagged particle and fluid particles is rather small.
At such a time scale, we should explicitly consider individual collision
events between the tagged particle and the surrounding fluid particles.
In this study, we analyzed the short-time dynamics of the tagged particle in an ideal gas,
where we do not have static nor hydrodynamic correlations between fluid particles. We performed event-driven hard sphere simulations and show
that the short-time dynamics of the tagged particle is correlated even
under such an idealized situation. 
Namely, the velocity autocorrelation function
becomes negative when the tagged particle
is relatively light and the fluid density is relatively high.
This result can be attributed to the dynamical
correlation between collision events.
To investigate the physical mechanism, 
which causes the dynamical correlation, we analyzed the
correlation between successive collision events. We found that
the tagged particle can collide with the same ideal gas particle several
times, and such collisions cause the strong dynamical correlation for
the velocity.
\end{abstract}

\maketitle

\section{introduction}

A particle immersed in a fluid exhibits 
a random motion, which is
widely known as the Brownian motion\cite{brown1828xxvii}. 
The Brownian motion originates
from the interaction between the tagged particle (tracer)
and the particles which compose the surrounding fluid (fluid particles).
In principle, the dynamics of the tracer is deterministic
because the full system, which consists of the tracer and the fluid
particles, obeys the Hamiltonian dynamics. However, if we observe only the
tracer, the dynamics looks stochastic (at least apparently).

For the description of such stochastic dynamics of the tracer,
the generalized Langevin equation (GLE)\cite{vanKampen2007,Sekimoto2010}
has been employed in many cases.
The GLE is a stochastic differential equation,
which incorporates the memory effect (memory kernel)
and the random noise. The properties of the memory kernel and
the random noise reflects the statistical properties of the surrounding fluid.
Formally, the projection operator method\cite{Kawasaki1973} 
gives the expression for the memory kernel and
the noise is related to the memory kernel via
the fluctuation-dissipation relation.
The dynamic equation for the tracer is given as
\begin{equation}
 \label{gle_general}
 M \frac{d^{2}\bm{r}(t)}{dt^{2}} = - \int_{-\infty}^{t} dt' \,
  K(t - t') \frac{d\bm{r}(t')}{dt'} + \bm{\xi}(t),
\end{equation}
where $\bm{r}$ and $M$ are the position and mass of the tracer, $K(t)$ is
the memory kernel, and $\bm{\xi}(t)$ is the random noise. The random noise
satisfies
\begin{equation}
 \label{gle_fdr}
 \langle \bm{\xi}(t) \rangle = 0, \qquad
 \langle \bm{\xi}(t) \bm{\xi}(t') \rangle = k_{B} T K(|t - t'|) \bm{1}, 
\end{equation}
where $\langle \dots \rangle$ represents the statistical average,
$k_{B}$ is the Boltzmann constant, $T$ is the temperature, and $\bm{1}$ is
the unit tensor.

The projection operator formalism does not tell us the full
properties of the noise $\bm{\xi}(t)$. It gives only the first and
second moments as eq~\eqref{gle_fdr}.
Therefore, in many practical cases, the noise is assumed to be Gaussian.
Then the dynamic equation for the tracer is fully specified and
can be analytically solved\cite{Fox1977}.
Such type of a GLE with the Gaussian noise
is utilized to analyze various experimental data\cite{mason1995optical,mason1997particle,li2010measurement,kheifets2014observation,huang2011direct}.
The diffusion of the tracer can be also directly studied by
molecular dynamics simulations.
Recently, the memory kernels for some systems are precisely
calculated from the molecular dynamics simulations\cite{Lesnicki2016,Straube2020}.
For example, the power-law type long-time tail, which originates from hydrodynamic modes,
is reported for a Lennard-Jones fluid\cite{Lesnicki2016}.
It should be noted that, in most cases, a tracer is assumed to be large and massive\cite{Petravic2008}.
However, in some cases, a tracer could be rather small and light. For example,
if we interpret a molecule which is dissolved into a fluid as a tracer,
the tracer size is comparable to the fluid particles and the mass of the
tracer is also comparable to that of a fluid particle.
(For example, if we consider a mixture system which consists of isotopes and
interpret a light species as a tracer, the tracer mass can be slightly 
smaller (or larger) than unity\cite{kiriushcheva2005influence}.)
The dynamics of
a tracer particle can be experimentally measured by such as the scattering,
and we need to describe the dynamics of the tracer particle to analyze
the experimental data.

Here we emphasize that the Gaussian noise approximation
is not justified {\it a priori}. Naively, we expect that
the approximation is reasonable because the noise would be
Gaussian as a results of the accumulated many forces due to the collisions of
surrounding fluid particles. As far as the number of collisions is
sufficiently large, the average force will converge to a Gaussian noise 
by the central limit theorem.
Conversely, when the number of collisions is not 
enough, the Gaussian noise approximation does not work properly.
Such a situation can be realized when we consider the dynamics
of a single tagged particle in a dilute gas. The Gaussianity of the noise
can be evaluated via the Gaussianity of the displacement, since
the displacement is a linear combination of the noise (by eq~\eqref{gle_general}).
For instance, Yamaguchi and Kimura\cite{yamaguchi2001non} investigated
the dynamics of a single hard-sphere in a dilute hard-sphere gas.
They reported that the distribution function for the displacement is
non-Gaussian at a short time scale.
We should be careful when we employ the Gaussian noise approximation.

The tracer dynamics at the short time scale in a dilute fluid can
be described, for example, by the gas kinetic theory\cite{chapman1990mathematical}.
To illustrate the dynamics with a small number of collisions,
we should consider the individual collision events explicitly.
Then, we expect that the dynamics would be modeled by the sequence
of the collision events.
Burshtein and Krongauz\cite{Burshtein1995negative}
modeled the dynamics of a tracer in a hard-sphere fluid
as a sequence of the collision events. They assumed simple
statistics for the collision events (the statistical distributions
of the waiting time between successive collisions and the velocity change distribution are
decoupled, and given as simple hypothetical forms) and 
calculated the dynamical
quantities such as the velocity autocorrelation function (VAC). 
Their model can successfully
reproduce some dynamical properties.
However, we note that the employed statistics
would be a matter of further discussion.
We expect that the statistics depends on the fluid density rather strongly.
When the fluid is dilute, then the collisions are statistically
almost independent\cite{chapman1990mathematical}. Some assumptions in
the Burshtein-Krongauz model will be justified in such a case.
However, when the fluid is rather dense, the situation becomes very complex.
In a dense hard-sphere fluid\cite{alder1970decay,herman1972studies}, 
the statistics of collisions
would depend on various factors. 
For example, it can be related to the structure of the fluid.
Recently,
Mizuta et al\cite{Mizuta2019} performed a series of molecular dynamics simulations
for a fullerene particle immersed in a liquid Argon.
They reported that the dynamics of fullerene particles is affected both by
the hydrodynamics and structure of the fluid.
Although their result is interesting, it seems difficult to
quantitatively separate the contributions of individual factors.

The statistics of collision events can also depend on the mass of
the tracer. 
For a collision between two hard spheres, 
velocities of spheres after the collision depend on the
velocities before the collision and masses of two hard spheres. 
The collision statistics depends on the cross-section 
of the collision and the relative velocity. For the collision
between a tracer particle and a fluid particle, the statistics of
the relative velocity depends on the masses of the tracer and the fluid particle.
Also, the momentum exchange between the tracer and the ideal gas particle
depends on the mass ratio.
Then, if the mass of the tracer is changed, the collision statistics
would be changed (by the change of the relative velocity distribution).
If the tracer mass is sufficiently larger than the fluid molecule mass, 
the momentum of the tracer will not
be largely affected by a single collision.
However, if the tracer mass is
relatively small, the momentum largely changes even by a single collision, and consequently 
the back reflection and the negative velocity autocorrelation of the tracer can occur\cite{alder1974studies,herman1972studies}.
The statistics of the collision
events would not be simple in this case.
Such collision statistics can be naturally handled by utilizing the
methods developped in the gas kinetic theory\cite{chapman1990mathematical,eder1977velocity}, instead of the GLE.
We can describe the dynamics by the Boltzmann equation or by the Fokker-Planck equation,
and then employ some collision statistics to calculate the dynamical quantities such
as the velocity autocorrelation.
For dilute gases, we may reasonably assume that collisions are statistically independent.
Then the collision distribution can be described by the Poisson process,
some dynamical quantities behave qualitatively different from the GLE with the Gaussian noise approximation.
For example, the mass dependence of the diffusion coefficient can be
explained by the gas kinetic theory.
If the gas density is relatively high, however, the description by the gas kinetic
theory becomes nontrivial. The Burshtein-Krongauz model would be interpreted
as a phenomenological extension of the kinetic theory to a relatively dense system.
The extension of the GLE by combining non-Gaussian type processes which mimic
collisions has also been proposed\cite{Gelin2014,gelin2015microscopic}.
However, the microscopic origin of the negative velocity correlation has not
been fully clarified yet.

In this paper, we investigated the dynamics of the tracer particle
at the short-time scale, from the view point of collision events.
As we mentioned, the statistics of collision events and the short-time dynamics
of the tracer depend both on the fluid density and the tracer mass. They
are also affected by several different factors such as the fluid structure
and the hydrodynamic interaction.
To eliminate such factors other than the fluid density and the tracer mass,
we consider systems where fluid particles do not interact with each other
(the ideal gas).
We investigated the dynamics of the tracer in the ideal gas by
the hard-sphere simulations. We performed series of simulations with
various parameter sets.
We found that the tracer
exhibits a rather complex dynamics at the short-time scale,
even by such an idealized and simplified model.
We analyzed the dynamics
of the tracer on the basis of the collision type dynamics
by Burshtein and Krongauz\cite{Burshtein1995negative}. 
Below, We will show that the correlation between sequential collisions
is important to describe the short-time dynamics of the tracer.

\section{Model and Method}

We consider the dynamics of a tracer immersed in an ideal gas 
by a numerical simulation.
In this work, we use the term ``an ideal gas'' as a gas composed 
of point masses, which do not interact with each other at all. The point masses
do not exchange their momenta via collisions.
The tracer collides with the ideal gas particles
whereas the ideal gas particles do not collide each other.
We modeled the tracer as a hard sphere particle, and
employed the standard hard sphere simulation method\cite{alder1959studies}.
Both the tracer and ideal gas particles move ballistically until they collide. 
The velocity of the tracer and ideal gas particles 
are instantaneously changed when they collide.

We consider a single tracer particle and $N$ ideal gas particles
in the cubic simulation box with the periodic boundary conditions.
We set the number of ideal gas particles as $N = 10^{5}$ for all the simulations.
We express the masses of the tracer and ideal gas particles as $M$ and $m$,
the size of the tracer as $\sigma$ (the size of the ideal gas particles is $0$),
the temperature of the system as $T$, and the number density of the ideal gas particles as $\rho$. (Here, the particle density is related 
to the system volume $\mathcal{V}$ as $\rho = N / (\mathcal{V}- \pi \sigma^ 3/6)$.)
We employ dimensionless units where the unit energy, unit mass, and unit length 
are $k_{B} T$ ($k_{B}$ is the Boltzmann constant), $m$ and $\sigma$, respectively. In the dimensionless units, the system can be 
characterized only by two parameters: $M$ and $\rho$.
We varied the density $\rho$ in the range from $10^{-2}$ to $10^{3}$,
and the mass $M$ in the range from $10^{-2}$ to $10^{3}$.

The initial state of the simulation was generated as follows.
The tracer particle was located at the box center, and the ideal
gas particles were dispersed randomly in the box. (The position of
a newly generated ideal gas particle
should not overlap to the tracer.)
The initial velocities of the tracer and ideal gas particles were
sampled from the Maxwell-Boltzmann distribution.
To prevent the center of mass of the system from drifting,
we subtracted the velocity of the center of mass from
all the particles in the system.
Then, the velocities of particles were rescaled to
reduce the average kinetic energy to $3/2$.
Since the momentum is conserved in the
hard sphere simulation, the center of mass does not move during a simulation.
After the initial state was generated, we evolved the system by
using the established procedure for the hard-sphere simulation\cite{alder1959studies}.

Although the hard-sphere simulation itself is rather simple and
clear, we should be careful about the handling of the images due to the
periodic boundary condition.
In this simulation, the mean free path of the tracer is long and this
leads to unexpected overlaps between the tracer and images of ideal gas particles.
To avoid such overlaps, we performed simulations as follows.
\begin{enumerate}
 \item Generate the initial state.
 \item \label{collision_detection_step}
       Find the ideal gas particle(the target particle), which collides the tracer
       judging from their positions and velocities.
       Let the waiting time for 
       the collision of this particle is $\Delta t_{1}$.
 \item \label{maximum_velocity_detection_step}
 	Find the ideal gas particle, which has the maximum relative velocity to the tracer
	along the axis of the simulation box. From the maximum relative velocity 
	$V_{\text{max}}$, obtain the minimum time of collision of the image particle to the
	tracer as $\Delta t_2 = (L-\sigma)/(2V_{\text{max}})$, 
       where $L(=\mathcal{V}^{1/3})$ is the box length.
 \item If $\Delta t_2 < \Delta t_1$, update the position of all particles by
       the step size $ \Delta t_2$, then proceed to $t \to t + \Delta t_{2}$
       and repeat the update unless $\Delta t_1 \le \Delta t_2$. 
       When $\Delta t_1 \le \Delta t_2$ is satisfied, update the position of all particles
       by the step size $\Delta t_1$, then proceed to $t \to t + \Delta t_{1}$
       to attain the collision.
 \item \label{velocity_update}
 	Update the velocity of the tracer and the target particle according to the collision.
 \item Repeat from step\ref{collision_detection_step} to step\ref{velocity_update}
 	until 10 million collisions.
\end{enumerate}

From the trajectories of the tracer, we calculated several
dynamical quantities: We calculate the mean-square displacement (MSD),
the non-Gaussian parameter (NGP)\cite{rahman1964correlation}, and the velocity autocorrelation function (VAC) defined as
\begin{equation}
 g(t) \equiv \langle \Delta \bm{r}^{2}(t) \rangle, 
\end{equation}
\begin{equation}
\alpha(t) \equiv \frac{3 \langle \Delta \bm{r}^{4}(t) \rangle}
  {5 \langle \Delta \bm{r}^2(t) \rangle^{2}} - 1,
\end{equation}
\begin{equation}
 C(t) \equiv \frac{\langle \bm{V}(t) \cdot \bm{V}(0) \rangle}
  {\langle \bm{V}^{2} \rangle} .
\end{equation}
Here, $\Delta \bm{r}(t) \equiv \bm{r}(t) - \bm{r}(0)$ is the displacement of the tracer
, and $\bm{V}(t)$ is the velocity of the tracer.
$\langle \dots \rangle$ represents the statistical
average.
We also calculated a few quantities which characterize the
change of the tracer velocity by collisions. They will be introduced later.

Before we show the simulation results, here we briefly comment
on the effect of the system size to the simulation results.
In our model, the ideal gas particles can change their velocities only via the collision
to the tracer. Thus the relaxation time for the velocity of an ideal gas particle
becomes very long and it depends on the system size. One may consider that
the velocity distribution of the ideal gas particles deviate from the
equilibrium distribution. Also, one may suspect that the dynamics of the tracer
strongly depends on the system size. As far as we examined, fortunately,
the velocities of the ideal gas particles obey the equilibrium Maxwell-Boltzmann
distribution. Some dynamical quantities of the tracer, such as the MSD and VAC,
do not show measurable system size dependence (unless the system size is
too small and comparable to the tracer size).
Therefore, we conclude that the system size dependence for simulation
results can be safely neglected (at least for the analyses shown below).

\section{Results}

Figure~\ref{fig:msd_d1_change_m}(a) shows the mass dependence of the MSD
for the ideal gas density of $\rho = 1$.
In a short-time region, the ballistic behavior
$\langle \Delta \bm{r}^{2}(t) \rangle \propto t^{2}$ is observed, and the 
MSD decreases as $M$ increases.
This behavior is consistent with the fact that the
average absolute value of velocity decreases as $M$ increases.
In a long-time region, diffusive behavior
$\langle \Delta \bm{r}^{2}(t) \rangle \propto t$ is observed, and the 
MSD decreases as $M$ increases.
This result means that the diffusion coefficient of tracer depends on $M$.
One may argue that the $M$-dependence is counter-intuitive 
because the GLE~\eqref{gle_general}
predicts that the inertia term does not contribute to the long-time dynamics.
However, from the view point of the gas kinetic theory, the diffusion coefficient can
depend on the mass\cite{chapman1990mathematical}.
A similar behavior has been reported for the motion of a tracer in Lennard-Jones fluids\cite{RN61,RN62}
and hard-sphere fluids\cite{alder1974studies}.
Nevertheless, in this work, we do not discuss the $M$-dependence of the diffusion
coefficient.

Figure~\ref{fig:msd_d1_change_m}(b) shows the $M$ dependence of the NGP.
The parameters are the same as Figure~\ref{fig:msd_d1_change_m}(a).
The NGP increases with time in the ballistic regime, exhibits a peak in the 
transitional regime, and decays in the diffusive regime.
The peak increases as $M$ decreases.
This result means that the statistics of the displacement is
non-Gaussian except in the diffusive regime when $M$ is small.
The non-Gaussian behavior is observed in various systems and often attributed to
the heterogeneity of the environment. For example, the glass forming
liquids\cite{kob1997dynamical} and polymer solutions\cite{xue2016probing} work as heterogeneous environments for a tracer.
However, in our system, the ideal gas particles never have the structures
and thus the non-Gaussian behavior cannot be attributed to the
heterogeneity of the environment.
Therefore, we consider that non-Gaussian behavior has
the kinetic origin. In our system, the tracer dynamics is affected
only by the collisions and the non-Gaussian behavior can be attributed to
the properties of collisions.
(Yamaguchi and and Kimura also reported similar non-Gaussian behavior for a dilute
hard sphere fluid\cite{yamaguchi2001non}.)

\begin{figure}[hbt]
 \begin{center}
\includegraphics[width=0.5\linewidth]{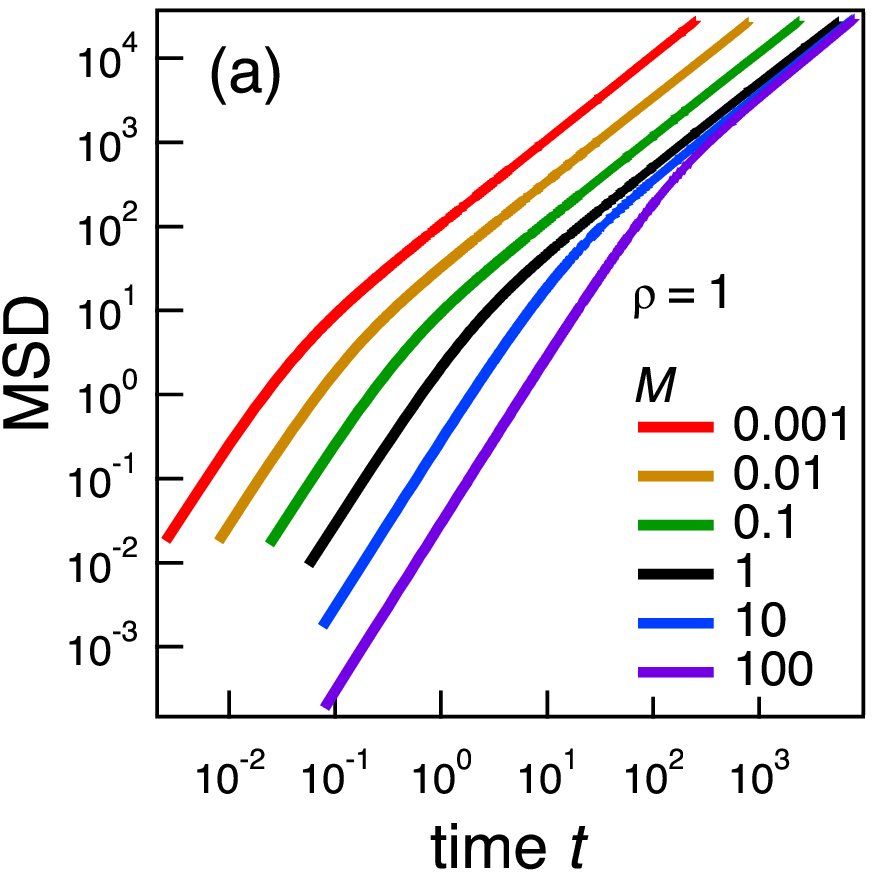}%
\includegraphics[width=0.5\linewidth]{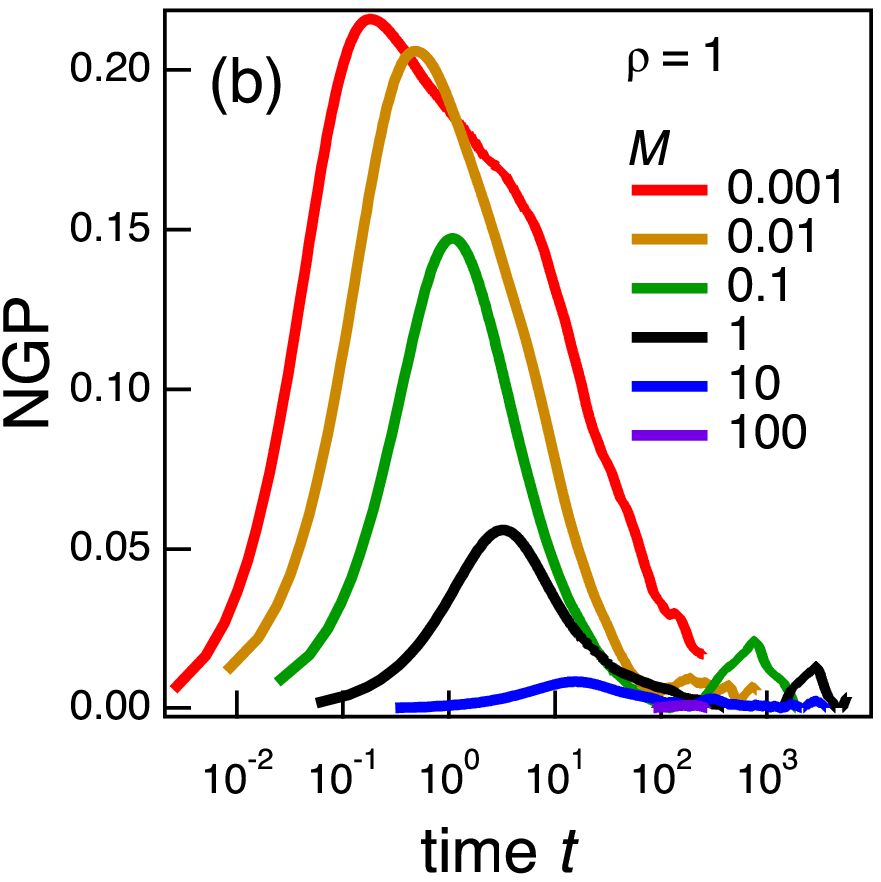}
 \end{center}
 \caption{(a) MSD and (b) NGP of a tracer in the ideal gas for various $M$.
 The density of the ideal gas is $\rho = 1$.}\label{fig:msd_d1_change_m}
\end{figure}

\begin{figure}[hbt]
 \begin{center}
\includegraphics[width=0.5\linewidth]{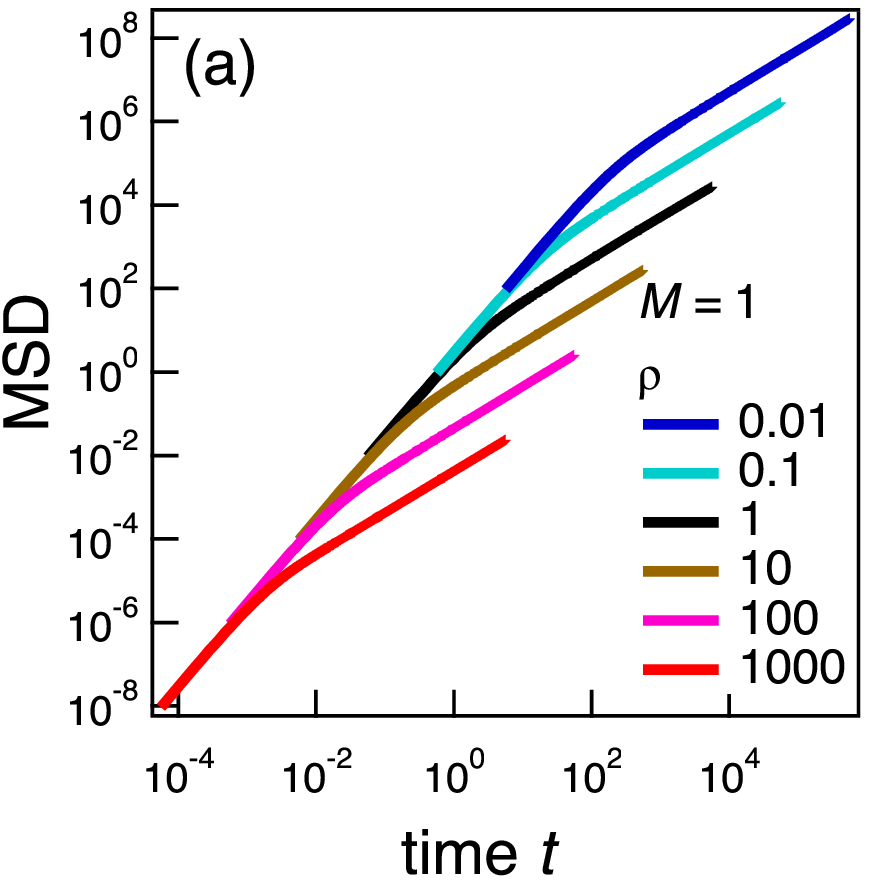}%
\includegraphics[width=0.5\linewidth]{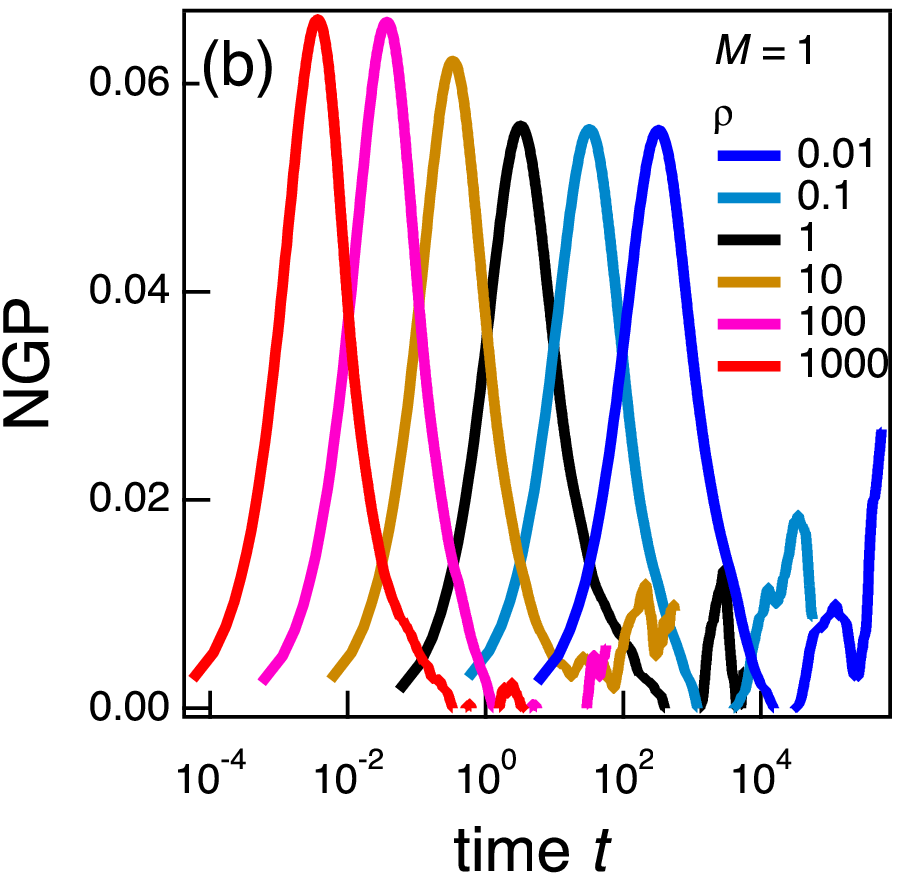}
 \end{center}
 \caption{(a) MSD and (b) NGP of a tracer in the ideal gas for various $\rho$. 
 The mass of the tracer is $M = 1$.}
 \label{fig:msd_m1_change_d}
\end{figure}

Figures~\ref{fig:msd_m1_change_d} shows the $\rho$-dependence of MSD
and NGP with a constant mass of the tracer ($M=1$).
We observe the MSD collapses into a single curve in the ballistic region.
The MSD deviates from the ballistic behavior to the diffusive behavior,
and the crossover time decreases as $\rho$ increases.
This $\rho$-dependence is due to the change of the mean free path.
As seen in Fig~\ref{fig:msd_d1_change_m}, 
we observe the non-negligible peak for the NGP.
The peak value of the NGP is almost constant for $\rho\le1$,
whereas the value slightly increases as $\rho$ increases for $\rho\ge1$.
Although such increase of NGP against $\rho$ is commonly
observed for liquids with heterogeneous structures\cite{van2014molecular}, 
we have no heterogeneity in our model, as mentioned above.

From the mass and density dependence shown above, we conclude that
the dynamics of the tracer qualitatively depends on the mass and the
density. If the mass is large ($M \gtrsim 1$), the dynamics of the particle
can be reasonably described as the Gaussian process.
(The GLE with the Gaussian noise approximation is reasonable.) If the density
is low ($\rho \lesssim 1$), the dynamics of the tracer can be 
expressed by the ballistic motion and collisions (as the standard gas kinetic theory\cite{chapman1990mathematical}).
However, if the mass is small and the density is high 
($M \lesssim 1$ and $\rho \gtrsim 1$), the dynamics
seems not to be expressed as a simple and intuitive model.
We consider that this is caused by the ``dynamic'' correlation.
(In our system, the fluid has no static structure since it is an ideal gas.)
Our result suggests that, the dynamic correlation solely gives such an
apparently counter-intuitive behavior. A simple interpretation is that
something like a ``cage'' would be dynamically formed and the tracer
effectively feels confinement at the short-time scale.
However, this behavior is not trivial, and to investigate it in detail,
we performed some additional
simulations for several parameter sets for $M \lesssim 1$ and $\rho \gtrsim 1$.

\begin{figure}[htb]
 \begin{center}
\includegraphics[width=0.5\linewidth]{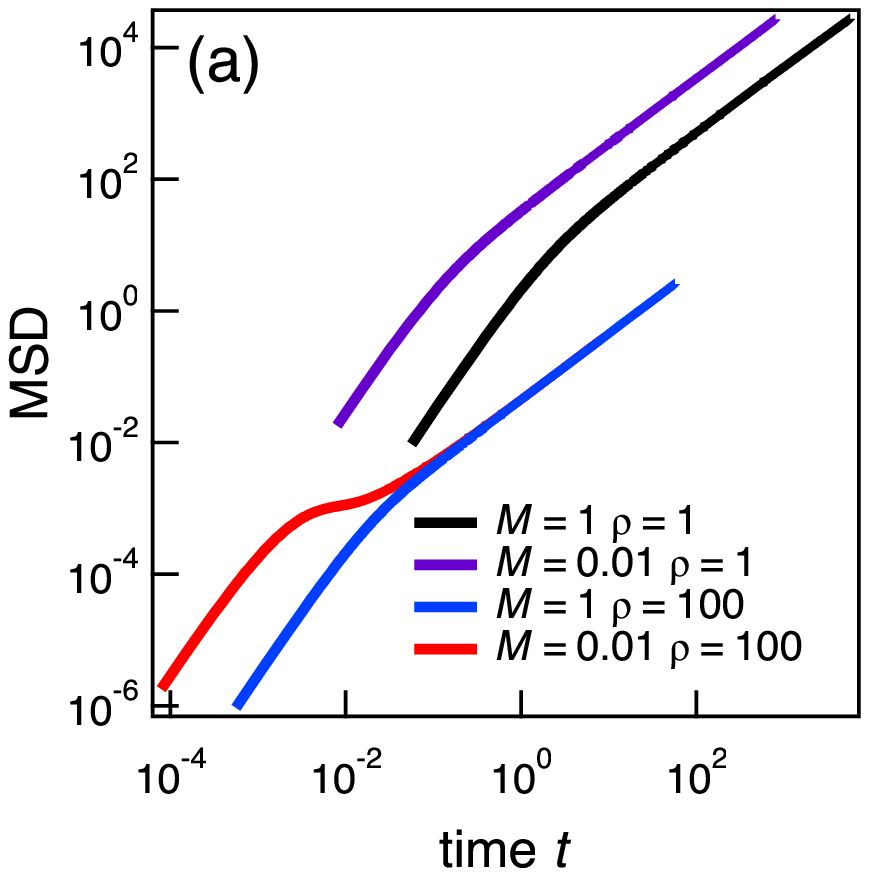}%
\includegraphics[width=0.5\linewidth]{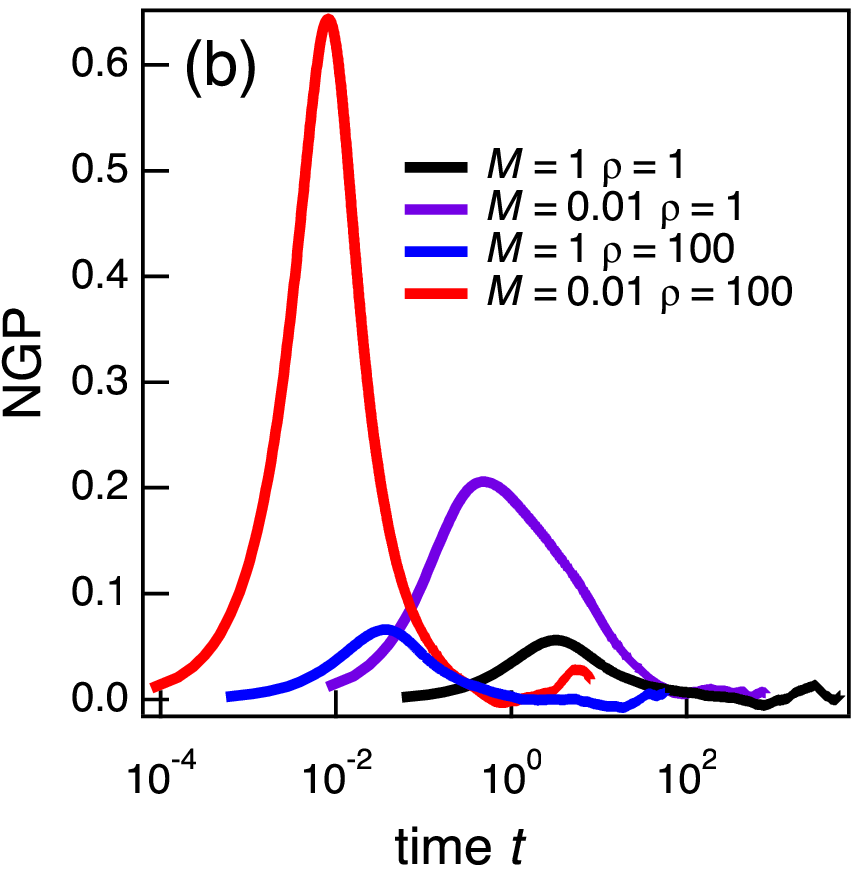}
 \end{center}
 \caption{(a)MSD and (b)NGP of a tracer for several sets
 of the mass $M$  and the ideal gas density $\rho$.
 \label{fig:msd_change_m_d} }
\end{figure}

Figure~\ref{fig:msd_change_m_d} shows the MSD and NGP for several different
parameter sets. From Figure~\ref{fig:msd_change_m_d}(a), we find that the MSD
shows unexpected behavior for $M = 0.01$ and $\rho = 100$.
Namely, the MSD does not show the plain crossover from the ballistic to diffusion
behavior, and we observe an intermediate subdiffusive
region between the ballistic and
diffusive regions.
The similar behavior of the MSD has been reported for
some systems such as glass forming liquids and polymer solutions\cite{RN65,RN66}. Although the data are not explicitly presented,
similar behavior has been also reported for the multi particle
collision dynamics (MPCD) type model\cite{Malevanets2000}.
In Figure~\ref{fig:msd_change_m_d}(b), 
we observe a strong peak of NGP for $M = 0.01$ and $\rho = 100$.
(Intuitively, the peak time for the NGP ($t_{\text{peak}}$) corresponds to
the relaxation time of the tracer velocity. This interpretation seems to be
consistent with the MSD data. The MSD exhibits the ballistic behavior 
for $t \lesssim t_{\text{peak}}$, and the diffusive behavior for $t \gtrsim t_{\text{peak}}$.)
Thus demonstrated, we confirm that our system exhibits
nontrivial dynamical behavior for
small $M$ and high $\rho$. These results seem not to be
simply described by the GLE or the gas kinetic theory.
Some questions naturally arises.
What is the origin of such nontrivial behavior? How can we
describe the dynamics?

\begin{figure}[htb]
 \begin{center}
  \includegraphics[width=0.5\linewidth]{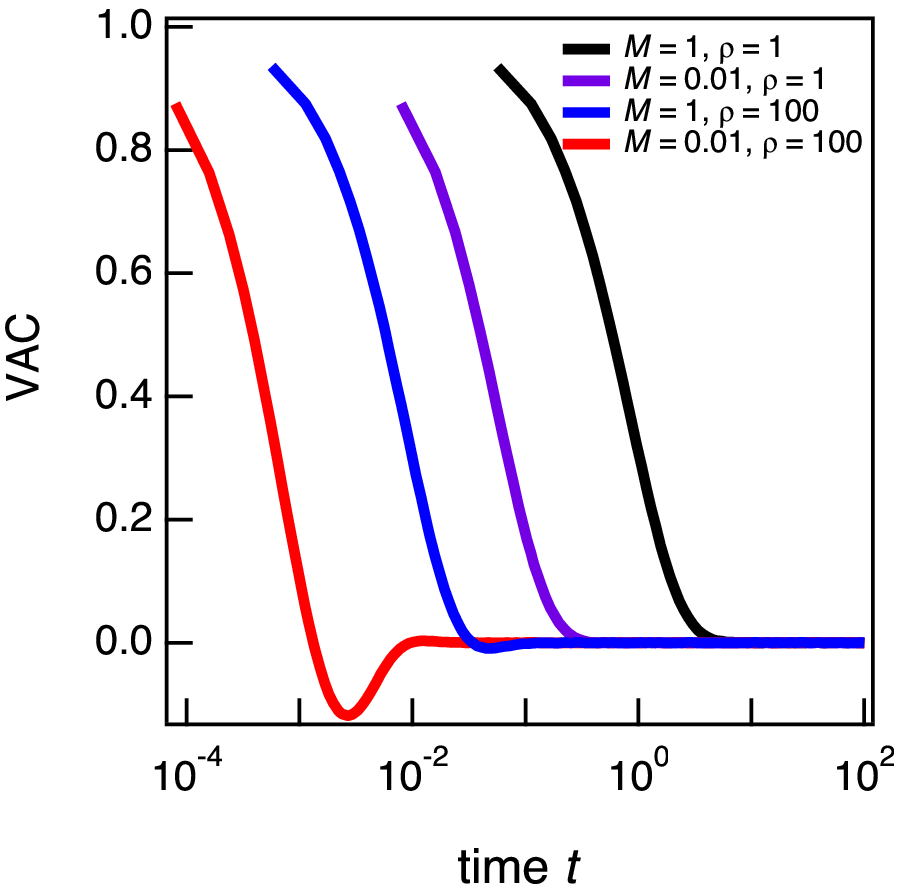}
 \end{center}
 \caption{VAC of a tracer for several sets of
 the mass and the ideal gas density. The parameters are the same as in
 Figure~\ref{fig:msd_change_m_d}.
 \label{fig:vacf_change_m_d}}
\end{figure}

Here we recall that our simulation model is a
collision-driven hard-sphere molecular dynamics model.
In our model, the tracer moves ballistically and the velocity
changes only by the collisions with the ideal gas particles. Therefore,
it would be reasonable for us to concentrate on the velocity of the
tracer. Figure~\ref{fig:vacf_change_m_d} shows the VAC of a tracer
particle for the same parameter sets as Figures~\ref{fig:msd_change_m_d}.
For the case of $M = 0.01$ and $\rho = 100$, the VAC
becomes negative at the intermediate time region.
The negative correlation in the VAC means that the tracer is
effectively reflected back by collisions (the back reflection).
A negative VAC is usually attributed to the strong static and dynamic
correlations between fluid particles. However, in our system,
the ideal gas particles do not have static and dynamic correlations
and thus we cannot attribute the negative VAC to the correlations between
fluid particles.
A possible intuitive interpretation is as follows.
If the mass of the tracer is small, the tracer has a large velocity compared
with the surrounding ideal gas particles.
Also, due to the large mass contrast, the momentum of an ideal gas particle
is almost not affected when the tracer and the ideal gas particle collides.
Thus, we may interpret the ideal gas particles as almost fixed obstacles.
Under such a situation, the back reflection behavior may occur.
Meanwhile, we should notice that the effect of the back reflection becomes especially
strong when the density is high.
To clarify the short-time dynamics,
therefore, we should analyze the collisions in detail.
In addition to the short-time dynamics, the long-time dynamics is
important for some analyses. The VAC typically exhibits so-called the long-time tail
at the long time region. In our simulation results, however, the power-law type
long-time tail is not clearly observed, at least in the
examined time scale. (We may observe the long-time tail at the very long
time scale, but it is beyond the scope of this work and thus we do not consider
the long time region in what follows.)

Burshtein and Krongauz\cite{Burshtein1995negative}
considered a similar negative velocity correlation
in hard sphere fluids. They analyzed the dynamics of a hard sphere carefully
and proposed a simple model which describes the motion of a hard sphere.
They modeled a process as a sequence of collision events, which we may
interpret as a renewal process. They assumed
that the time interval between two successive collisions is sampled from a
waiting time distribution $P(\tau)$, and there is no correlation
between successive waiting times. They also assumed that the velocity of the hard
sphere is stochastically changed from $\bm{V}$ to $\bm{V}'$, and
the new velocity $\bm{V}'$ is sampled from the velocity distribution $P(\bm{V}'|\bm{V})$. Their model claims that the negative velocity
correlation can emerge if both of the following two conditions are
satisfied; (a) the waiting time distribution $P(\tau)$ is
not a exponential distribution,
and (b) the average velocity after a collision is negatively correlated
to the velocity before the collision, $\langle \bm{V}' \cdot \bm{V} \rangle < 0$. Although their model seems plausible, some assumptions in their
model cannot be fully justified. Thus we investigate the statistics of collisions
in our simulations below.

\begin{figure}[htb]
 \begin{center}
  \includegraphics[width=0.5\linewidth]{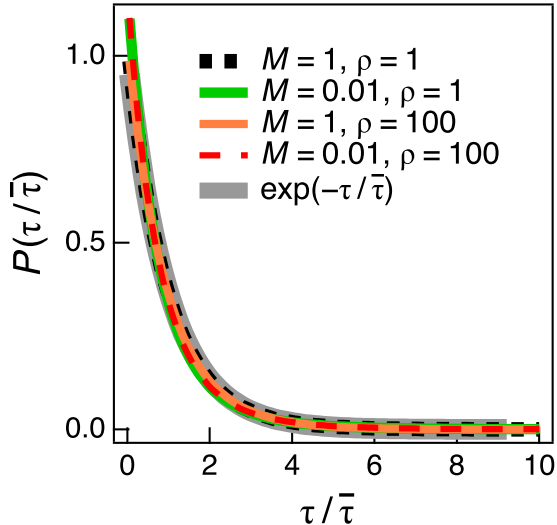}
 \end{center}
 \caption{The waiting time distribution function $P(\tau)$
 with different values of the mass $M$ and the ideal gas density $\rho$.
 The waiting time $\tau$ is normalized by the average $\bar{\tau}$ as
 $\tau / \bar{\tau}$. For comparison, the exponential distribution
 $P(\tau) = (1 / \bar{\tau}) \exp(- \tau / \bar{\tau})$ is also plotted. 
 \label{fig:tau_dis_change_m_d}}
\end{figure}

Figure~\ref{fig:tau_dis_change_m_d} shows the waiting time distribution
between two successive collision events. 
In this work, we define the waiting time as the time interval between
two successive collision events. (We do not distinguish whether the tracer collides to
the same ideal gas particle or not.)
The waiting time $\tau$ 
is normalized by the average waiting time $\bar{\tau} \equiv \int d\tau
\, \tau P(\tau)$. We observe that the distribution function for the
normalized waiting time is almost independent of $M$ and $\rho$.
Moreover, the waiting time distribution can be reasonably fit to the
exponential distribution.
Naively, the exponential waiting time distribution means that
the collision events are statistically independent. Thus this result
implies that the Poisson distribution for the number of collisions
during a finite time period. However, this naive argument is physically not
fully justified. This is because the statistics of the waiting time generally
depends on the tracer velocity.
The observed waiting time distribution should be interpreted as the
average waiting time distribution over the tracer velocity.
This average waiting time distribution seems to be well described by
the exponential distribution.
(We show the calculations for the waiting time distributions in Appendix~\ref{waiting_time_distribution}.)
Anyway, the assumption (a) is not
satisfied.
According to the Burshtein-Krongauz model, if the 
waiting time distribution is exponential, then the VAC is always positive.
Clearly, this is not consistent with our simulation result in
Figure~\ref{fig:vacf_change_m_d}.
We expect that some assumptions by Burshtein and Krongauz would not
be valid (at least for our system). We will discuss the statistical
properties of collisions in detail, in the next section.

\section{Discussions}

\subsection{Gaussianity}

In this subsection, we discuss the Gaussianity of the dynamics 
of the tracer. Our simulation results
(Figures~\ref{fig:msd_d1_change_m}-\ref{fig:msd_change_m_d})
show that the tracer exhibits the non-Gaussian behavior even if the
surrounding fluid is an ideal gas. The NGP strongly depends on the
tracer mass $M$, and if $M$ is sufficiently large, the NGP becomes small.
Therefore, if $M$ is sufficiently large, the
dynamics of the tracer can be reasonably described by a Gaussian
process. For such a case, the GLE with the Gaussian noise can be
utilized.

As we mentioned, the GLE with the Gaussian noise
has been widely employed to analyze experimental
data. But our results imply that the application of the GLE may not be
fully justified in some cases. Here we discuss validity of the GLE
for some experimental systems.
Recent progress of experimental techniques enabled us to
experimentally measure the short-time ballistic motion of
a tracer particle\cite{li2010measurement,huang2011direct}.
Li et al\cite{li2010measurement} investigated the dynamics of $3\text{{\textmu}m}$ beads
in dilute gases ($2.75 \rm{kPa}$ and $99.8\rm{kPa}$) at the micro second scale.
The mass of the tracer is about $10^{-10} \text{g}$, whereas
the mass of the gas molecule is about $10^{-23} \text{g}$.
Clearly the mass ratio is very large (about $10^{13}$)
and thus we expect that the tracer dynamics is reasonably described by the GLE.
Note that the memory effect is negligible in this system and thus the GLE reduces to a simple
Langevin equation.
Huang et al\cite{huang2011direct} measured the dynamics of a tracer in a fluid.
Although the fluid density is much larger than the experiments by Li et al,
the mass ratio is estimated to be the same order as that of Li et al.
Therefore, we conclude that for typical experimental conditions for the
short-time dynamics of a tracer particle, the GLE (or the Langevin equation) with the Gaussian noise
can be safely employed.

However, the argument above does not fully justify the use of the Langevin
equation to analyze the other experimental data. The Langevin equation is
utilized for various analyses of experimental data, even at the very
small objects. Some examples are the light absorbance in the infrared (IR) spectroscopy
and the diffusion in the nuclear magnetic resonance (NMR).
In the IR spectroscopy, the absorbance is often calculated based on the
oscillator model for polar functional groups. A functional group is
not sufficiently small compared with the surrounding objects, and the
analyses based on the Langevin equation may not be justified.
Actually, Roy et al\cite{Roy2011} measured two dimensional correlation
spectra and reported that the dynamics is not Gaussian.
In the NMR, the diffusion of a proton (or other atom) is estimated from
the NMR signal based on a simple diffusion equation by the Brownian
motion. (Even if the Langevin equation is not formally justified,
we empirically know the analyses based on the simple
Langevin equation gives reasonable results in most cases.)
We consider that if the conventional analyses do not give 
physically reasonable result, we should analyze the data carefully
based on the collision dynamics.
(Strictly speaking, we should employ quantum dynamics to analyze
the IR or NMR data\cite{Tanimura2006}. However, the naive incorporation
of the Gaussian noise like a Langevin thermostat to quantum systems
is questionable\cite{VanKampen2005}. Some ideas in this work might
be useful to consider the modeling of thermostats.)

\subsection{Correlation between Collisions}

In this subsection, we consider the collision events in detail.
As we showed, the waiting time distribution $P(\tau)$ in our system
is well described by the exponential distribution.
 (The assumption (a) in
the Burshtein-Krongauz model is not satisfied.)
The Burshtein-Krongauz model does not give a negative VAC for the exponential waiting time distribution.
Here, it should be mentioned that
the exponential waiting time distribution is also reported for
hard sphere fluids\cite{talbot1992statistical}.
The reason why the waiting time obeys the exponential distribution is
not clear. Because our system is similar to hard sphere systems in some
aspects, we expect there is a common mechanism which gives
the exponential waiting time distribution.

In the Burshtein-Krongauz model, the correlation between
the velocities before and after the collision is also an important quantity.
To investigate the statistical
properties of the collision events in detail,
we consider the correlation between velocities by introducing the quantity
defined as 
$\gamma_{1} =
\langle \bm{V}' \cdot \bm{V} \rangle_{\text{coll}} / \langle \bm{V}^2 \rangle_{\text{coll}}$
(where $\bm{V}$ and $\bm{V}'$
are the velocities of the tracer before and after a collision event and 
$\langle \cdots \rangle_{\text{coll}}$ represents the statistical average with respect to collisions).
If we assume that the distribution of the velocity after the collision is Gaussian,
a collision event statistics can be characterized only by this $\gamma_{1}$.
In the Burshtein-Krongauz model, $\gamma_{1}$ should be negative to give
a negative VAC.

\begin{figure}[htb]
 \begin{center}
  \includegraphics[width=0.5\linewidth]{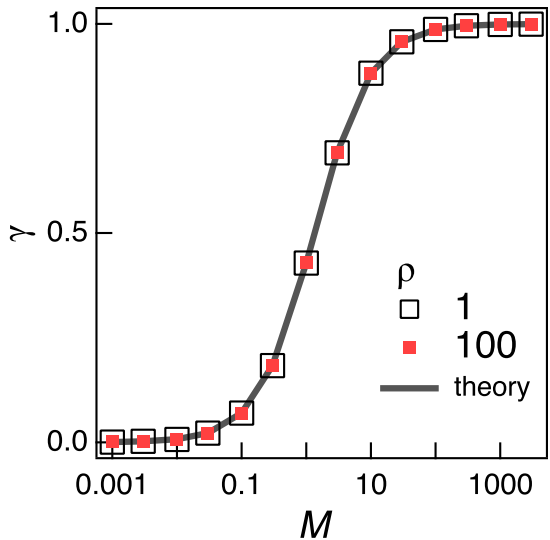}
 \end{center}
 \caption{The correlation factor of the velocity before
 and after a collision, $\gamma_{1}$. The mass $M$ dependence of $\gamma_{1}$ with two different values of $\rho$ is shown. The data for different $\rho$
 are almost the same and data points look fully overlapped.
 \label{fig:gamma1_change_m_d}}
\end{figure}

It would be informative to analyze $\gamma_{1}$.
In our system, the statistics of the fluid particles is simple and thus we can theoretically calculate $\gamma_{1}$, 
which can be expressed in a simple form as
\begin{equation}
 \label{gamma1_theoretical_model_reslt}
 \gamma_{1} = \frac{3M}{3M+4} .
\end{equation}
See Appendix~\ref{calculation_of_gamma_1} for the details of the model and calculations. Here, we should
note that $\gamma_{1}$ given by eq~\eqref{gamma1_theoretical_model_reslt}
is independent of the ideal gas density $\rho$.
Therefore, we expect that $\gamma_{1} > 0$ holds for any $M$ and $\rho$.
We show the prediction by eq~\eqref{gamma1_theoretical_model_reslt}
together with the simulation data
in Figure~\ref{fig:gamma1_change_m_d}.
We find that eq~\eqref{gamma1_theoretical_model_reslt} reasonably
reproduces the simulation result without any fitting parameters.
From these results, we conclude that
the velocity correlation factor $\gamma_{1}$ cannot be negative
in our model. This means that the assumption (b) in the
Burshtein-Krongauz model is not satisfied, neither.
From these results,
both two assumptions in the Burshtein-Krongauz model are not 
satisfied in our system. However, the original idea of the Burshtein-Krongauz model that
the dynamics of the tracer is described by the successive
collision events seems to be reasonable. We expect that with several
modifications, the Burshtein-Krongauz type model may explain the
dynamical behavior. Before we consider the modification, we should
investigate the detailed statistical properties of the collision events.

\begin{figure}[hbt]
 \begin{center}
  \includegraphics[width=0.5\linewidth]{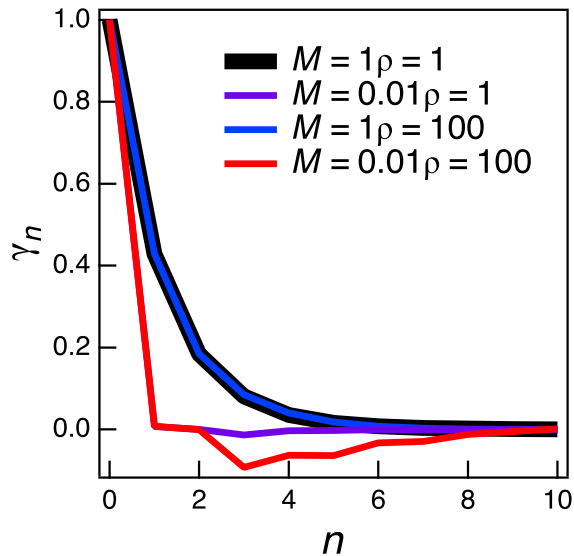}
 \end{center}
 \caption{The correlation factor of the velocity for multiple collisions, $\gamma_{n}$. $n$ represents the number of collisions.
 The data for different values of the mass $M$ and the ideal gas density $\rho$ are shown.
 \label{fig:gamma_num_change_m_d}}
\end{figure}

We consider that the successive collisions can be correlated in our
system. Then the correlation factor $\gamma_{1}$ is not sufficient to characterize the
properties of the collision events. We should investigate multiple
collisions which can be correlated.
We define the correlation factor for the multiple collision events as
\begin{equation}
 \gamma_{n} \equiv \frac{\langle \bm{V}_{i + n} \cdot \bm{V}_{i} \rangle_{\text{coll}}}{\langle \bm{V}_{i}^{2} \rangle_{\text{coll}}} .
\end{equation}
If the successive collision events are independent, we have $\gamma_{n} = \gamma_{1}^{n}$. If the successive collision events are not statistically
independent, $\gamma_{n}$ may behave in a different way.
Figure~\ref{fig:gamma_num_change_m_d} shows the correlation factor
$\gamma_{n}$ directly calculated from the simulations. For $M = 1$ and
$\rho = 1$, $\gamma_{n}$ decays exponentially as $n$ increases, which
seems to be consistent with the estimate based on the statistically independent
collisions. For $M = 0.01$ and
$\rho = 100$, however, we find that $\gamma_{n}$ does not exhibit the
exponential decay. Clearly, $\gamma_{n}$ has a negative peak at $n = 3$,
while $\gamma_{1}$ and $\gamma_{2}$ are almost zero. Therefore,
the successive collision events are not statistically independent
if the mass is small and the ideal gas density is high.
As a consequence, we cannot simply apply the Burshtein-Krongauz model.

Based on this result, here we attempt to
construct the modified version of the Burshtein-Krongauz model.
From the simulation results of the correlation factor data
for the multiple collisions $\gamma_{n}$, it would be possible to treat
several collision events (for example, $n = 3$ or $4$ events) as a coarse-grained event. Then we can express the correlation factor
for the coarse-grained collision event as $\gamma_{\text{eff}} = \gamma_{n}$.
Also, the waiting time distribution between two coarse-grained
events becomes
\begin{equation}
 \label{coarse_grained_waiting_time_distribution}
 \begin{split}
  P_{\text{eff}}(\tau)
  & = \int d\tau_{1} \dots d\tau_{n} \, \delta(\tau - \tau_{1} - \dots \tau_{n})
  P(\tau_{1}) \dots
  P(\tau_{n}) \\
  & = \frac{t^{n - 1}}{(n - 1)! \tau^{n}}   \exp(-\tau / \bar{\tau}) .
 \end{split}
\end{equation}
Eq~\eqref{coarse_grained_waiting_time_distribution} is the gamma
distribution, and is clearly different from the exponential
distribution, for $n \ge 2$.
If we assume that coarse-grained collision events are statistically
independent, we can utilize the Burshtein-Krongauz model by interpreting
their collision events as the coarse-grained collision events.
With such a coarse-graining,
the negative velocity correlation can be reasonably explained by
the (modified) Burshtein-Krongauz model.

\subsection{Origin of Negative Correlation}

\begin{figure}[htb]
 \begin{center}
  \includegraphics[width=0.5\linewidth]{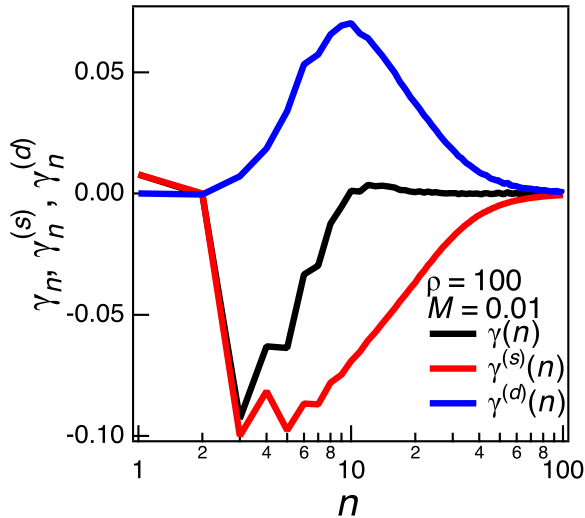}
 \end{center}
 \caption{The self and distinct part of the correlation factor
 for multiple collisions, $\gamma_{n}^{(s)}$ and $\gamma_{n}^{(d)}$.
 The mass and the ideal gas densities are $M = 0.01$ and $\rho = 100$.
 The full correlation factor is given as $\gamma_{n} = \gamma_{n}^{(s)} + \gamma_{n}^{(d)}$.
 \label{fig:gamma_num_sd}}
\end{figure}

Although the Burshtein-Krongauz model seems to work with some
modifications, it does not explain why the correlation factor $\gamma_{n}$ becomes negative for $n = 3$. To clarify the emergence mechanism of the
negative correlation, we analyze the correlation factor in further detail.
Because now we know that the successive collisions are not independent,
it would be reasonable to consider whether the correlation comes from
the same ideal gas particle or not. It would be informative
to decompose the correlation factor $\gamma_{n}$ into the correlation
by the collisions to the same ideal gas particle and the collisions to
different ideal gas particles. We thus define the self and distinct
parts of the correlation factor for multiple collisions:
\begin{align}
 \gamma_{n} & = \gamma_{n}^{(s)} + \gamma_{n}^{(d)}, \\
 \gamma_{n}^{(s)} & \equiv \frac{\langle \bm{V}_{i + n} \bm{V}_{i}
 \, I(K_{i + 1} = K_{i + n}) \rangle} {\langle \bm{V}_{i}^{2} \rangle} , \\
 \gamma_{n}^{(d)} & \equiv 
 \frac{\langle \bm{V}_{i + n} \bm{V}_{i}
 \, I(K_{i + 1} \neq K_{i + n}) \rangle}{\langle \bm{V}_{i}^{2} \rangle} ,
\end{align}
where $I(\dots)$ takes $1$ if the argument is true and takes $0$ if not,
and $K_{i}$ represents the index of the ideal gas particle which collides
to the tracer at the $i$-th collision event.
Figure~\ref{fig:gamma_num_sd} shows the self and distinct parts of the
collision factor, for $M = 0.01$ and $\rho = 100$. From
Figure~\ref{fig:gamma_num_sd}, we find that $\gamma_{n}^{(s)}$ is negative for $n \ge 3$
whereas $\gamma_{n}^{(d)}$ is positive for all $n$.
This result means that the collisions by different particles do not
contribute to the negative correlation, and the negative correlation
emerges from the collisions with the same particle.

The reason why the correlation occurs at
$n = 3$ can be now understood rather intuitively. We consider the
sequence of several collision events. At the first collision ($n = 1$),
the correlation is always positive ($\gamma_{1} > 0$), as we demonstrated.
By this first collision, the tracer and the ideal gas particle exchange
the momentum and they move apart from each other. Therefore at the
second collision ($n = 2$) the tracer cannot collide to the same ideal
gas particle as the first collision. After the second collision, the
momentum of the tracer is changed again, and the tracer can move towards the
ideal gas particle which collided at the first collision. Thus, for $n \ge 3$,
we have the correlation with the same ideal gas particle.

Before we end this subsection, we briefly mention about
the MSD. As known well, the MSD and the VAC can be related via the Kubo formula\cite{kubo1966fluctuation}. The MSD is obtained by doubly integrating the VAC. Conversely,
we have the velocity correlation as
\begin{equation}
\frac{d^2\langle{\Delta \bm{r}}^2(t)\rangle}{dt^2}=\langle {\bm V}(t)\cdot {\bm V
}(0) \rangle .
\end{equation}
If the velocity correlation becomes negative, at a certain time we
should have $\langle \bm{V}(t)\cdot \bm{V}(0) \rangle = 0$.
This fact
means that if we have the negative velocity correlation, we may
observe an inflection point in the MSD. In addition, if the negative
correlation is strong, the MSD may exhibit an intermediate region
between the short-time ballistic and the long-time diffusive regions.
Indeed, we observed such nontrivial intermediate region in Figure~\ref{fig:msd_change_m_d}(a). Interestingly, the time scale where
we observe this intermediate region seems to be the same where we
observe the strong peak of the NGP in Figure~\ref{fig:msd_change_m_d}(b).
Therefore, we consider that the collision with the same ideal gas particle
is the origin of various nontrivial short-time behavior of the tracer.

\subsection{Comparison with Other Simulations}

In this subsection, we compare our results with the other simulations.
Mizuta et al\cite{Mizuta2019} investigated the dynamics of the single fullerene in the liquid Argon by
using the molecular dynamics simulation. 
They investigated the dynamics of the fullerenes at various time scales.
At the short time scale where the time is shorter than the Enskog time,
they showed that the Enskog theory\cite{chapman1990mathematical} 
reasonably describes the VAC of a fullerene.
The Enskog theory is based on the gas kinetic theory, and thus the
results by Mizuta et al implies that the short-time scale dynamics is governed
by collisions.
The Enskog time is the time scale where approximately only one collision per fluid particle occurs.
At the longer time scales, the behavior deviates from the Enskog theory.
Mizuta et al reported the effect of the fullerene size. For the small
fullerene (or for a fluid particle), they reported that the VAC becomes negative
at the short-time scale region. This result is qualitatively consistent with
ours. However, we should stress that fluid particles strongly interact with each other in their system.
Therefore, we cannot quantitatively compare our result with their simulation data.
We expect that our analyses would be useful to study some aspects of the
short-time scale dynamics in realistic systems like those investigated by Mizuta et al.

At the long-time scale where the time is
longer than the kinematic time of the momentum diffusion over the particle
size, they reported that the GLE works reasonably.
Intuitively, at the long-time scale, the fluid can be treated as a continuum
fluid rather than discrete particles. Then the GLE can be safely employed.
This intuitive expectation consistent with their result. Conversely,
our system does not have the interactions between the fluid particles,
and thus it seems not reasonable to treat fluid particles as
a continuum fluid. In our system,
the fluid particles move ballistically until they collide with the tracer.
Because we employed the periodic boundary condition,
the mean free path $\lambda$ of the ideal gas particles depends on the system size
as $\lambda \approx {\mathcal{V}} / {\sigma^2}$.
The Knudsen number $\textrm{Kn}$ can be estimated as follows:
\begin{equation}
 \mathrm{Kn} =\frac{\lambda}{\sigma}=\frac{\mathcal{V}}{\sigma^3}.
\end{equation}
Clearly, the obtained $\mathrm{Kn}$ is much larger than $1$
for sufficiently a large system ($\mathrm{Kn} \gg 1$).
Therefore, we conclude that
the ideal gas can not be regarded as a continuum fluid for the tracer.
Consequently, our system does not obey usual continuum hydrodynamic
equations such as the Navier-Stokes equation.
Our system is designed to study the short-time dynamics and should not
be used to study the long-time dynamics.
This would explain why we do not observe the long-time tail
in the long time region. To apply the continuum hydrodynamic equations to
our model, we should consider very long time scale and very large length
scale. The long-time tail may be observed at such a very long time scale,
although it would be longer than the current simulation time scale.
Because the fluid cannot be expressed by usual continuum hydrodynamic
equations, it would not be straightforward to interpret the
non-Markovian memory effect in our system. (Normally, the memory kernel can
be related to the molecular scale hydrodynamics\cite{Lesnicki2016,Straube2020,Mizuta2019}.)
As far as the author knows, it is not clear whether the collision-based description can be
successfully related to the hydrodynamic description or not.
It would be an interesting future work to study how the collision-based and
hydrodynamic descriptions are connected.

We also compare our results with the simulation by Alder and Alley\cite{alder1974studies}.
They investigated the effect of the tracer mass and the tracer size on the
dynamics in a hard sphere fluid.
They analyzed the VAC of the tracer 
and showed that the VAC of the tracer has the negative peak when the mass of the
tracer is small.
This result is consistent with our result.
According to our analyses, such a negative VAC can be attributed to 
the property of the multiple (typically three) sequential collisions.
Although the VAC data by Alder and Alley are not analyzed in terms of
collisions, their data seems to be consistent with our analyses:
the VAC becomes negative at the time where three collisions occur on average.
This implies that our theory holds, at least qualitatively, even under
the existence of the hard sphere interaction between fluid particles.
We expect that the interpretations and analyses based on the collision
events are useful to study the short-time dynamics of various systems.

\section{Conclusions}

We investigated the short-time dynamics of the tracer
in the ideal gas.
Our system can be characterized only by two parameters, the mass of
the tracer $M$ and the fluid density $\rho$, in the dimensionless units.
We performed hard-sphere type simulations and calculated various
dynamical quantities such as the MSD, NGP, and VAC, for various values of
$M$ and $\rho$.
The dynamical behavior is largely affected by $M$ and $\rho$.
For $M \gtrsim 1$, the NGP is very small
and the GLE with the Gaussian noise can describe the dynamics well.
For $M \lesssim 1$, however, the dynamics cannot be described
by a Gaussian stochastic process.
We should explicitly consider collision events to describe the dynamics
for $M \lesssim 1$.
For $\rho \lesssim 1$, the behavior is relatively simple and seems to be
consistent with the standard gas kinetic theory.
Meanwhile, for $M \lesssim 1$ and $\rho \gtrsim 1$,
some nontrivial dynamical
properties have been observed. For example, the VAC exhibits a negative value.

To study the origin of such non-trivial behavior, we analyzed the
collision events in detail.
We showed that the waiting time distribution between
two successive collisions is almost the exponential distribution,
and the correlation factor $\gamma_{1}$ is almost positive.
The Burshtein-Krongauz model
predicts that the VAC is always positive for our system, but it is not
consistent with the simulation result.
We introduced the correlation factor between
multiple collisions, $\gamma_{n}$, to further investigate
the statistical properties of collision events.
We showed that the velocities before and after multiple
collisions are rather strongly correlated, and this correlation is the
origin of the negative VAC. In addition, we showed that the strong
correlation mainly comes from the multiple collisions between the
tracer and the same ideal gas particle.

Although our system is highly idealized and simplified, we believe
that our results are informative to analyze realistic systems.
The dynamical correlation between multiple collision events would
exist in realistic systems. The analyses based on the collision picture,
such as the analysis of the correlation factor $\gamma_{n}$ and the
decomposition of it to $\gamma_{n}^{(s)}$ and $\gamma_{n}^{(d)}$ would
be useful to analyze the short-time dynamics in various systems.
The analyses for the systems with interactions such as
the hard sphere fluid and the Lennard-Jones fluid would be an
interesting future work.

\appendix

\section{Waiting Time Distribution}
\label{waiting_time_distribution}

In this appendix, we roughly estimate the waiting time distirbution
for the tracer particle in an ideal gas.
The collision statistics depends on the relative velocity of the tracer to
an ideal gas particle. The distribution of the relative velocity is generally not that simple
(as we consider in Appendix~\ref{calculation_of_gamma_1}).
However, under some special conditions, the distribution reduces to rather simple forms.
If the tracer mass is sufficiently large ($M \gg 1$), the tracer velocity becomes very
small compared with the surrounding ideal gas particles.
The relative velocity distribution
approximately coincides to the Maxwell-Boltzmann distribution of the
ideal gas particle. Then, the collision statistics do not depend on the
tracer velocity $\bm{V}$. Individual collision events are expected to be statistically
independent, and then the number of collisions during a finite time period
should obey the Poisson distribution. This Poission distribution can be
directly related to the exponential waiting time distribution.
Therefore, we expect that the waiting time distribution becomes exponential
if $M$ is sufficiently large.

On the other hand, if the tracer mass is sufficiently small ($M \ll 1$),
the relative velocity distribution is almost the same as the tracer
velocity $\bm{V}$. Here we consider the case where the tracer velocity
$\bm{V}$ is fixed to be constant. In this case, the relative velocity distribution
is also fixed and we will observe several collision events during a finite
time period. The individual collision events are statistically independent,
and thus we expect the exponential waiting time distribution. Therefore,
we have
\begin{equation}
 P(\tau|\bm{V}) = \frac{1}{\bar{\tau}(\bm{V})} \exp(- \tau / \bar{\tau}(\bm{V})) ,
\end{equation}
where $\bar{\tau}(\bm{V})$ is the average waiting time for the tracer velocity $\bm{V}$.
We expect that the interval between collisions decreases as the tracer velocity increases.
Naively, we consider the average waiting time can be related to the tracer velocity as
$\bar{\tau}(\bm{V}) = \Lambda / |\bm{V}|$ where $\Lambda$ is a mean free path,
and we assume $\Lambda$ is independent of $\bm{V}$. (The mean free path $\Lambda$ depends on the ideal gas density $\rho$.)
Then we can express
the waiting time distribution as
\begin{equation}
 \label{waiting_time_distribution_large_mass_limit}
 P(\tau) = \int d\bm{V} P(\tau|\bm{V}) P_{\text{eq}}(\bm{V})
  = \int_{0}^{\infty} dV \, \frac{V}{\Lambda} \exp(- \tau V / \Lambda) P_{\text{eq}}(V),
\end{equation}
where $V \equiv |\bm{V}|$, and $P_{\text{eq}}(\bm{V})$ and $P_{\text{eq}}(V)$
represent the equilibrium distribution functions for $\bm{V}$ and $V$, respectively.
The waiting time distribution can be then calculated to be
\begin{equation}
 \label{waiting_time_distribution_large_mass_limit_explicit}
 \begin{split}
  P(\tau) 
  & = \int_{0}^{\infty} dV \, \frac{4 \pi V^{3}}{\Lambda} 
  \left( \frac{M}{2 \pi}\right)^{3/2}
  \exp\left(- \frac{\tau V}{\Lambda} - \frac{M V^{2}}{2} \right) \\
  & = 
 \frac{1}{\Lambda \sqrt{M}} 
  \left[ - (3 + \tilde{\tau}^{2}) \tilde{\tau}
  \exp\left(\frac{\tilde{\tau}^{2}}{2} \right)
  \mathrm{erfc}\left( \frac{\tilde{\tau}}{\sqrt{2}} \right)
  + \sqrt{\frac{2}{\pi}} (2 + \tilde{\tau}^{2})
  \right],
 \end{split}
\end{equation}
with $\tilde{\tau} \equiv \tau / \Lambda \sqrt{M}$.
Eq~\eqref{waiting_time_distribution_large_mass_limit_explicit} can be reasonably 
approximated by the exponential distribution:
\begin{equation}
 P(\tau) \approx  \frac{1}{\Lambda \sqrt{M}} 
 \sqrt{\frac{8}{\pi}} \exp\left( - \sqrt{\frac{8}{\pi}} \frac{\tau}{\Lambda \sqrt{M}} \right).
\end{equation}
Thus if $M$ is sufficiently small, the waiting time distribution
approximately reduces to the exponential distribution with the average waiting
time $\bar{\tau} \approx \Lambda \sqrt{\pi  M / 8}$.

From the estimates shown above, both for sufficiently large and sufficiently small
 $M$ cases, we have the exponential distributions.
At the intermediate $M$ cases, we intuitively expect
that the waiting time distribution would be expressed in a similar expression
to eq~\eqref{waiting_time_distribution_large_mass_limit_explicit}.
We replace $V$ by the relative velocity $v' = |\bm{V} - \bm{v}|$
($\bm{v}$ is the velocity of an ideal gas particle):
\begin{equation}
 \label{waiting_time_distribution_general}
 P(\tau) 
  = \int_{0}^{\infty} dv' \, \frac{v'}{\Lambda} \exp(- \tau v' / \Lambda) P_{\text{eq}}(v'),
\end{equation}
where $P_{\text{eq}}(v')$ is the equilibrium relative velocity distribution.
The relative velocity obeys the Maxwell-Boltzmann distribution and thus
we approximately have the exponential waiting time distribution
(as the case of the sufficiently small $M$ case).
Therefore the waiting time distribution will
be well approximated by the exponential distribution for any $M$.
This is consistent with the simulation results shown in Figure~\ref{fig:tau_dis_change_m_d}.

\section{Calculation of $\gamma_{1}$}
\label{calculation_of_gamma_1}

In this appendix, we show the detailed calculation of
the correlation factor $\gamma_{1}$. We consider the equilibrium
system, in which the ideal gas particles are distributed uniformly in space,
and the velocity is sampled from the Maxwell-Boltzmann distribution.

In what follows, we calculate $\gamma_{1}$ in the dimensional units (not in
the dimensionless units in the main text) for the sake of clarity.
We describe the velocities before and after the collision as
$\bm{V}$ and $\bm{V}'$. We describe the velocity of the ideal gas
particle which collides to the tracer as $\bm{v}$. We calculate $\gamma_{1} =
\langle \bm{V}' \cdot \bm{V} \rangle_{\text{coll}} / \langle \bm{V}^2 \rangle_{\text{coll}}$ under
the following assumptions:
\begin{enumerate}
 \item The ideal gas particles are uniformly distributed in space.
 \item Both the tracer and ideal gas particles obey
       the Maxwell-Boltzmann distribution: \label{distribution}
 \begin{align}
P_{t}(\bm{V}) & = \left( \frac{M}{2\pi k_BT} \right)^{3/2} \exp\left( -\frac{M\bm{V}^2
 }{2k_BT}\right),  \\
 P_{g}(\bm{v}) & = \left( \frac{m}{2\pi k_BT} \right)^{3/2} \exp\left( -\frac{m\bm{v}^2
 }{2k_BT}\right).
 \end{align}
\end{enumerate}

We consider the velocity of the tracer after the collision with
an ideal gas particle. The velocity after the collision $\bm{V}'$ is
simply obtained from the momentum and the energy conservation:

\begin{equation}
\bm{V}' = \bm{V} + \frac{2m}{M+m}\hat{\bm{r}} \cdot
 ( \bm{v} - \bm{V}) \hat{\bm{r}} ,
\label{eq:ap_conservation0}
\end{equation}
where $\hat{\bm{r}}$ is the unit vector between the tracer and the ideal gas particle when the collision occurs (the direction vector).
To calculate $\gamma_{1}$, we need to calculate the ensemble
average. We can express the average of the internal product of velocities $\langle \bm{V}' \cdot \bm{V} \rangle_{\text{coll}} $ as
\begin{equation}
\langle \bm{V}' \cdot \bm{V} \rangle_{\text{coll}} = 
\frac{
\int d\bm{V} d\bm{v} d\hat{\bm{r}} \, 
P_{t}(\bm{V})  P_{g}(\bm{v})  P_{d}(\hat{\bm{r}}) 
| \bm{v} - \bm{V}| \,
\bm{V}' \cdot \bm{V}
 }{
\int d\bm{V} d\bm{v} d\hat{\bm{r}} \, 
P_{t}(\bm{V})  P_{g}(\bm{v})  P_{d}(\hat{\bm{r}}) 
| \bm{v} - \bm{V}|
 } ,
\label{eq:ap_conservation1}
\end{equation}
where $P_{d}(\hat{\bm{r}})$ represents the equilibrium distribution
of the direction vector. The factor $|\bm{v} - \bm{V}|$ comes from the fact
that the collision frequency is proportional to the relative velocity
between the tracer and the ideal gas particle. (The average with
respect to collision $\langle \dots \rangle_{\text{coll}}$ is different from
the simple equilibrium average by this factor.)
From eqs~\eqref{eq:ap_conservation0} and \eqref{eq:ap_conservation1},
\begin{equation}
\gamma_1 = 1+ \frac{2m}{M+m}
\frac{
\int d\bm{V} d\bm{v} d\hat{\bm{r}} \,
P_{t}(\bm{V})  P_{g}(\bm{v})  P_{d}(\hat{\bm{r}})
\left| \bm{v} - \bm{V}\right|
 \, \hat{\bm{r}} \cdot (  \bm{v} - \bm{V}) \hat{\bm{r}} \cdot \bm{V}
 }{
\int d\bm{V} d\bm{v} d\hat{\bm{r}} \,
P_{t}(\bm{V})  P_{g}(\bm{v})  P_{d}(\hat{\bm{r}})
\left| \bm{v} - \bm{V}\right|
 \bm{V}^2
 } .
\label{eq:ap_conservation2}
\end{equation}
We can rewrite eq~\eqref{eq:ap_conservation2} in a simpler form
by introducing the reduced mass and the relative velocity.
We define
$\nu = M + m$, $\mu = M m / \nu$,
$\bm{u} = (M\bm{V} + m\bm{v})/\nu$, 
$\bm{v}' =\bm{v} -\bm{V}$.
Then eq~\eqref{eq:ap_conservation2} reduces to
\begin{equation}
\gamma_1 = 1+
\frac{2m}{M+m}\frac{\displaystyle
\int  d\bm{u} d\bm{v}' d\hat{\bm{r}} \, P_{d}(\hat{\bm{r}}) |\bm{v}'|
( \hat{\bm{r}} \cdot \bm{v}' ) \,
 \hat{\bm{r}} \cdot
 \left( \bm{u}- \frac{m}{\nu} \bm{v}' \right)
 \exp\left(
-\frac{\nu\bm{u}^2+\mu\bm{v}'^2}{2k_BT}
\right)
 }{\displaystyle
\int  d\bm{u} d\bm{v}' d\hat{\bm{r}} \,P_{d}(\hat{\bm{r}}) |\bm{v}'|
 \left( \bm{u}- \frac{m}{\nu} \bm{v}' \right)^{2}
\exp\left(
-\frac{\nu\bm{u}^2+\mu\bm{v}'^2}{2k_BT}
\right)
}
\label{eq:ap_conservation3}
\end{equation}

We calculate the integral in the denominator of the second term in
the right hand side of eq~\eqref{eq:ap_conservation3}. The integrand does
not depend on $\hat{\bm{r}}$ and thus we can easily perform the integral
over $\hat{\bm{r}}$. Then we have
\begin{equation}
 \begin{split}
  & \int  d\bm{u} d\bm{v}'  d\hat{\bm{r}} \,P_{d}(\hat{\bm{r}}) |\bm{v}'| 
 \left( \bm{u}- \frac{m}{\nu} \bm{v}' \right)^{2}
\exp\left(
-\frac{\nu\bm{u}^2+\mu\bm{v}'^2}{2k_BT}
\right) \\
  & = \int d\bm{u} d\bm{v}' |\bm{v}'| 
 \left( \bm{u}- \frac{m}{\nu} \bm{v}' \right)^{2}
\exp\left(
-\frac{\nu\bm{u}^2+\mu\bm{v}'^2}{2k_BT}
\right) \\
  & = \pi^{3/2}\left(2k_BT\right)^{9/2}
\left( \frac{3}{\nu^{5/2}\mu^{2}}
+ \frac{4m^2}{\nu^{7/2}\mu^{3}} \right) .
\label{eq:denominator1}
 \end{split}
\end{equation}
Next we calculate the integral in the numerator of
the second term in the right hand side of eq~\eqref{eq:ap_conservation3}.
Without loss of generality, we can set the direction of the relative velocity
parallel to the $z$-axis: $\bm{v}' = [0, 0, v']$. For convenience,
we rewrite the direction vector in the spherical coordinate, as
$\hat{\bm{r}}=[\sin\theta\cos\phi, \sin\theta\sin\phi, \cos\theta]$.
To proceed the calculation, we need the distribution function
for the direction vector. Now a collision occurs between the tracer
particle fixed at the origin and the ideal gas particles which moves
parallel to the $z$-axis, with the velocity $v'$.
Then the distribution of the normal vector should be proportional to
the direction vector which is projected onto the $xy$ plane:

\begin{equation}
P_{d}(\bm{\hat{r}}_n) \propto |\sin\theta\cos\theta| .
\end{equation}
In addition, not all the ideal gas particles can collide to
the tracer.
For $v'>0$, the probability distribution becomes
\begin{align}
P_{d}(\theta, \phi) =
 \begin{cases}
  -{\sin\theta\cos\theta}/{\pi} & (\pi/2 \le \theta \le \pi) ,\\
  0 & (\text{otherwise}) .
 \end{cases}
 \label{eq:prob_r_pos}
\end{align}
Similarly, for $v' < 0$, we have
\begin{align}
P_{d}(\theta, \phi) =
 \begin{cases}
  {\sin\theta\cos\theta}/{\pi} & (0 \le \theta \le \pi/2) ,\\
  0 & (\text{otherwise}) .
 \end{cases}
 \label{eq:prob_r_neg}
\end{align}

By utilizing eqs~\eqref{eq:prob_r_pos} and \eqref{eq:prob_r_neg},
finally we have
\begin{equation}
\begin{split}
   & \int  d\bm{u} d\bm{v}' d\hat{\bm{r}} \, P_{d}(\hat{\bm{r}}) |\bm{v}'|
 ( \hat{\bm{r}} \cdot \bm{v}' ) \,
 \hat{\bm{r}} \cdot
 \left( \bm{u}- \frac{m}{\nu} \bm{v}' \right)
 \exp\left(
 -\frac{\nu\bm{u}^2+\mu\bm{v}'^2}{2k_BT}
 \right) \\
  & = - 2\pi^{3/2}\left(2k_BT\right)^{9/2}\frac{m}{\nu^{5/2}\mu^3} .
 \label{eq:numetaror1}
\end{split}
\end{equation}
By substituting eqs~\eqref{eq:denominator1} and \eqref{eq:numetaror1}
into eq~\eqref{eq:ap_conservation3}, we have the following simple
expression for $\gamma_{1}$:
\begin{equation}
 \gamma_1
 = \frac{3M}{3M+4m} .
\end{equation}
This gives eq~\eqref{gamma1_theoretical_model_reslt} in the main text. From this derivation, it would be rather trivial that $\gamma_{1}$ does
not have the negative value.

\bibliographystyle{apsrev4-1}
\bibliography{references-20200817}
\end{document}